\documentstyle[multicol,prb,aps,ifthen,epsfig]{revtex}

\newcommand{\wide}[2]{                                                        %
\end{multicols}                                                               %
\widetext                                                                     %
\noindent                                                                     %
\ifthenelse{\equal{#1}{t}}                                                    %
{}                                                                            %
{                                                                             %
\raisebox{0.1in}[0in][0.02in]{$\rule{3.575in}{0.002in}                        %
\rule{0.002in}{0.08in}$}                                                      %
}                                                                             %
#2                                                                            %
\ifthenelse{\equal{#1}{b}}                                                    %
{}                                                                            %
{                                                                             %
{\raisebox{-0.1in}[0in][0.02in]                                               %
{\hspace{3.575in}$\rule{0.002in}{0.08in}                                      %
\rule[0.08in]{3.575in}{0.002in}$}                                             %
}                                                                             %
}                                                                             %
\begin{multicols}{2}                                                          %
\noindent                                                                     %
}                                                                             %

\def  \tG     {\tilde{G}}
\def  \tGo     {\tilde{G_0}}

\def  \tH     {\tilde{H}}
\def  \bsig    {\mbox{\boldmath$\sigma$}}
\def  \bssig   {\mbox{{\scriptsize \boldmath$\sigma$}}}
\def  \balph   {\mbox{\boldmath$\alpha$}}
\def  \bnab    {\mbox{\boldmath$\nabla$}}
\def  \bpi    {\mbox{\boldmath$\pi$}}
\def  \bspi   {\mbox{{\scriptsize \boldmath$\pi$}}}
\def  \eff     {\mathit{eff}}
\def  \Beff    {{\bf B}_{\mathit{eff}}}
\def  \op      {{\bf p}}
\def  \no      {\nonumber}

\begin{document}

\title{Relativistic corrections in magnetic systems}
\author{A. Cr\'{e}pieux and P. Bruno}
\address{Max-Planck-Institut f\"{u}r Mikrostrukturphysik,
Weinberg 2, 06120 Halle, Germany}
\date{\today}
\maketitle

\begin{abstract}
We present a weak relativistic limit comparison between the
Kohn-Sham-Dirac equation and its approximate form containing the
exchange coupling, which is used in almost all relativistic codes
of density functional theory. For these two descriptions, an exact
expression of the Dirac Green's function in terms of the
non-relativistic Green's function is first derived and then used
to calculate the effective Hamiltonian, i.e. Pauli Hamiltonian,
and effective velocity in the weak relativistic limit. We point
out that, besides neglecting orbital magnetism effects, the
approximate Kohn-Sham-Dirac equation also gives relativistic
corrections which differ from those of the exact Kohn-Sham-Dirac
equation. These differences have quite serious consequences: in
particular, the magneto\-crystalline anisotropy of an uniaxial
ferromagnet and the anisotropic magnetoresistance of a cubic
ferromagnet are found from the approximate Kohn-Sham-Dirac
equation to be of order $1/c^2$, whereas the correct results
obtained from the exact Kohn-Sham-Dirac equation are of order
$1/c^4$. We then give a qualitative estimate of the order of
magnitude of these spurious
terms. 
\end{abstract}

\begin{multicols}{2}

\section{Introduction}
The relativistic effects play a fundamental role in magnetic
systems. They are responsible for multiple physical properties of
great fundamental interest and technological
relevance\cite{Wilson91,Ebert96}: magnetic anisotropy;
magnetostriction; magneto-optical phenomena such as Faraday
effect, Kerr effect or magnetic dichroism; anomalous Hall effect
and anisotropic magnetoresistance (AMR) in metallic materials;
etc... To describe these properties, one has to use either a full
relativistic Hamiltonian, i.e. Dirac equation, or an effective
Hamiltonian, i.e. Pauli equation, which includes magnetic
interactions and spin-orbit coupling. However, it is not obvious
to find all the different terms which have to be taken into
account in the effective Hamiltonian. For this reason, it appears
necessary to study the weak-relativistic limit of the Dirac
equation for magnetic system in order to extract the complete
expression of the effective Hamiltonian.

For many-body systems, the relativistic density functional theory
allows to replace the many-body Dirac equation by the
Kohn-Sham-Dirac equation, which has the
form\cite{Strange98,Eschrig96,Dreizler90}
\begin{eqnarray}\label{DKS}
  H^A=c\balph\cdot\left(\op-e{\bf A}_{\eff}\right)+\beta mc^2+V_{\eff} ,
\end{eqnarray}
where $\balph$ and $\beta$ are the $(4 \times 4)$ Dirac matrices
respectively related to the $(2 \times 2)$ Pauli matrix $\bsig$
and unit matrix
\begin{equation}\label{dmat}
\balph\equiv\left(\begin{array}{clcr}
  0 & \bsig \\
  \bsig & 0
  \end{array}\right),
  \;\;\;\;\;\;\;\;\;\;\;\;\; \beta\equiv\left(\begin{array}{clcr}
  1 & 0 \\
  0 & -1
  \end{array}\right)  .
\end{equation}
The effective potential $V_{\eff}$ and effective vector potential
${\bf A}_{\eff}$ are functions of external and internal fields as
well as electron density $n({\bf r})=\sum_{i=1}^N \psi_i^+({\bf
r})\psi_i({\bf r})$ and current density ${\bf J}({\bf
r})=\sum_{i=1}^N \psi_i^+({\bf r})c\balph\psi_i({\bf r})$:
\begin{eqnarray}\label{effpot}
    V_{\eff}({\bf r})&\equiv&V_{\mathit{ext}}({\bf r})
    +\int\frac{n({\bf r}')}{|{\bf r}-{\bf r}'|}d{\bf r}'
    +\frac{\delta E_{\mathit{xc}}[n({\bf r}),{\bf J}({\bf r})]}{\delta n({\bf r})},\\
    {\bf A}_{\eff}({\bf r})&\equiv&{\bf A}_{\mathit{ext}}({\bf r})
    +\frac{e^2}{4\pi\epsilon_0c}\int\frac{{\bf J}({\bf r}')}{|{\bf r}-{\bf r}'|}d{\bf r}'\no\\
    &&+\frac{1}{e}\frac{\delta E_{\mathit{xc}}[n({\bf r}),{\bf J}({\bf r})]}{\delta {\bf J}({\bf r})}.
\end{eqnarray}
where $V_{\mathit{ext}}$ and ${\bf A}_{\mathit{ext}}$ are the
external potential and vector potential and $E_{\mathit{xc}}$ is
the exchange-correlation energy functional.

Neglecting orbital currents, a Gordon decomposition allows to
rewrite (\ref{DKS}) as\cite{Rajagopal73,MacDonald79}
\begin{eqnarray}\label{hrel}
  H^B=c\left(\balph\cdot\op\right)+\beta mc^2+V_{\eff}-\mu_B\beta\left(\bsig\cdot\Beff\right) .
\end{eqnarray}
The last term\cite{Note1} corresponds to the exchange coupling
involving an effective magnetic field $\Beff={\bf
B}_{\mathit{ext}}+{\bf B}_{\mathit{xc}}$ where ${\bf
B}_{\mathit{ext}}$ is the external field and ${\bf
B}_{\mathit{xc}}$ the exchange-correlation field given by
\begin{eqnarray}\label{Bxc}
  {\bf B}_{\mathit{xc}}({\bf r})=\frac{1}{\mu_B}\frac{\delta E_{\mathit{xc}}[n({\bf r}),{\bf m}({\bf r})]}{\delta {\bf m}({\bf r})}.
\end{eqnarray}
where ${\bf m}({\bf r})=\sum_{i=1}^N \psi_i^+({\bf
r})\beta\bsig\psi_i({\bf r})$ is the spin density and $\mu_B
\equiv e\hbar/2m$. Expression (\ref{hrel}) for the Hamiltonian
can also be derived and justified with the help of symmetry
arguments\cite{Thaller92}.

To describe relativistic effects in magnetic systems, one could
either start from $H^A$ or from $H^B$. Even if the second
Hamiltonian is an approximate description in comparison to the
former one, it has two advantages: on the first hand, it is
simpler because the vector potential is not more present in the
kinetic energy and, on the other hand, it has a more convenient
form for magnetic system because we can include the magnetic
interactions directly in the scalar potential which is then a sum
of spin-independent and spin-dependent parts. Therefore, $H^B$ is
generally
used\cite{Feder83,Strange84,Ujfalussy96,Razee97,Kraft95,Ebert96a,Ebert96b}.
However, there is no check on the consistency of these two
equations in the weak relativistic limit. Apart from the orbital
magnetism which is clearly neglected in $H^B$, we should obtain
the same relativistic corrections. We point out in the present
paper that the weak relativistic limits of $H^A$ and $H^B$
differ. These differences have quite serious consequences: in
particular, the magnetocrystalline anisotropy of an uniaxial
ferromagnet and the AMR of a cubic ferromagnet are found from the
approximate form $H^B$ to be of order $1/c^2$, whereas the
correct results obtained from the exact form $H^A$ are of order
$1/c^4$. We then give a qualitative estimate of the order of
magnitude of these spurious terms.

The paper is organized in the following way. In Sec. \ref{DGF}, we
present the derivation of a useful expression of the Dirac Green's
function in term of the non-relativistic Green's function. Then,
using this Dirac Green's function as a alternative method to
separate particles from anti-particles, we extract in Sec.
\ref{WRL} the effective Hamiltonian and effective velocity. These
operations are done starting both from $H^A$ and $H^B$ in order
to compare the relativistic corrections of order $1/c^2$. The
calculations starting from $H^A$ are given in detail in the text
whereas the calculations from $H^B$, which are similar in their
principles but lead to different expressions, are summarize in the
Appendices. Finally in Sec. \ref{Conc}, we discuss the results
and present a qualitative evaluation of the difference between
the two descriptions.

\section{Dirac Green's function}\label{DGF}
The derivation of the Dirac Green's function has been done both
for $H^A$ and $H^B$. Below, we present in detail the calculation
for $H^A$ while the calculation for $H^B$ is summarize in the
Appendix A. In order to simplify the notations, we introduce
$\bpi\equiv\op-e{\bf A}_{\eff}$, and write simply $V$ instead of
$V_{\eff}$. Since we are interested mainly in the electrons states
in the weak relativistic limit, we shift the zero of energy by
$-mc^2$; for positrons, we would have to shift the zero of energy
by $+mc^2$. We express the Hamiltonian $H^A$ in terms of $(2
\times 2)$ matrices as:
\begin{eqnarray}\label{hrel2}
  H^A=\left(\begin{array}{cc}
  V & c\left(\bsig\cdot\bpi\right) \\
  c\left(\bsig\cdot\bpi\right) & V-2mc^2
  \end{array}\right) .
\end{eqnarray}
A series of algebraic manipulations allows to express the Dirac
Green's function $G(z) \equiv (z-H^A)^{-1}$ in terms of the
($2\times 2$) non-relativistic Green's function $\tG(z) \equiv (z-
\tH^A)^{-1}$ associated with the ($2\times 2$) non-relativistic
Hamiltonian $\tH^A \equiv (\bsig\cdot\bpi)^2 / 2m + V$. We write
first $G$ as a product of matrices:
\begin{eqnarray}\label{gint1}
  G(z)&=&
  \left(\left(\begin{array}{cc}
  z & -c\left(\bsig\cdot\bpi\right) \\
  -c\left(\bsig\cdot\bpi\right) & 2mc^2
  \end{array}\right)\right. \no \\
  & &+\left.\left(\begin{array}{cc}
  -V & 0 \\
  0 & z-V
  \end{array}\right)\right)^{-1} \no \\
  &=&\left(1+{\mathcal A}^{-1}(z)
  \left(\begin{array}{cc}
  -V & 0 \\
  0 & z-V
  \end{array}\right)\right)^{-1}
  {\mathcal A}^{-1}(z) ,
\end{eqnarray}
where we have introduced:
\begin{eqnarray}\label{A}
  {\mathcal A}(z)=\left(\begin{array}{cc}
  z & -c\left(\bsig\cdot\bpi\right) \\
  -c\left(\bsig\cdot\bpi\right) & 2mc^2
  \end{array}\right) .
\end{eqnarray}
The explicit inversion of ${\mathcal A}$ yields
\begin{eqnarray}\label{gint2}
  {\mathcal A}^{-1}(z)=\left(\begin{array}{cc}
  \tGo(z) & \frac{\bssig\cdot\bspi}{2mc}\tGo(z) \\
  \frac{\bssig\cdot\bspi}{2mc}\tGo(z) & \frac{z}{2mc^2}\tGo(z)
  \end{array}\right) ,
\end{eqnarray}
where $\tGo(z) \equiv (z-(\bsig\cdot\bpi)^2/2m)^{-1}$. Using
(\ref{gint2}) in (\ref{gint1}), we get
\begin{eqnarray}\label{gint3}
  G(z)&=&
  \left(\begin{array}{cc}
  1-\tGo(z)V & \frac{\bssig\cdot\bspi}{2mc}\tGo(z)(z-V) \\
  -\frac{\bssig\cdot\bspi}{2mc}\tGo(z)V & 1+\frac{z}{2mc^2}\tGo(z)(z-V)
  \end{array}\right)^{-1} \no \\
  & &\times\left(\begin{array}{cc}
  \tGo(z) & \frac{\bssig\cdot\bspi}{2mc}\tGo(z) \\
  \frac{\bssig\cdot\bspi}{2mc}\tGo(z) & \frac{z}{2mc^2}\tGo(z)
  \end{array}\right) .
\end{eqnarray}
The first factor in the right hand side of the above equation is calculated
by direct matrix inversion, and is equal to
\wide{m}{
\begin{eqnarray}\label{gint4}
  \left(\begin{array}{cc}
  \tG(z)\tGo^{-1}(z)-\tG(z)\frac{\bssig\cdot\bspi}{2mc}(z-V)D(z)\frac{\bssig\cdot\bspi}{2mc}
  \left(\tG(z)\tGo^{-1}(z)-1\right) & -\tG(z)\frac{\bssig\cdot\bspi}{2mc}(z-V)D(z) \\
  D(z)\frac{\bssig\cdot\bspi}{2mc}\left(\tG(z)\tGo^{-1}(z)-1\right) & D(z)
  \end{array}\right) ,
\end{eqnarray}
where we have used the relation $\tGo V \tG=\tG-\tGo$ and
introduced the $(2\times 2)$ matrix
\begin{eqnarray}\label{expD}
  D(z) &=& \left(1+Q(z)\frac{(z-V)}{2mc^2}\right)^{-1} ,
\end{eqnarray}
with $Q(z)=1+(\bsig\cdot\bpi)\tG(z)(\bsig\cdot\bpi)/2m$. We report
(\ref{gint4}) in (\ref{gint3}) and finally obtain:
\begin{eqnarray}\label{green2}
  G(z)=\left(\begin{array}{cc}
  \tG(z)-\tG(z)\frac{\bssig\cdot\bspi}{2mc}
        (z-V)D(z)\frac{\bssig\cdot\bspi}{2mc}\tG(z)
  & \tG(z)\frac{\bssig\cdot\bspi}{2mc} Q^{-1}(z)D(z)Q(z)\\
  D(z)\frac{\bssig\cdot\bspi}{2mc}\tG(z)
  & \frac{1}{2mc^2}D(z)Q(z)
  \end{array}\right).
\end{eqnarray}
}
The corresponding expression associated with $H^B$ is given in the
Appendix A by Eq.~(\ref{greenB1}). We have thus succeeded in
expressing the ($4\times 4$) Dirac Green's function in terms of
($2\times 2$) matrices, in particular the non-relativistic
Green's function $\tG$. This formulation provides an alternative
method to separate particles form anti-particles in the
weak-relativistic limit by a simple block diagonalization. More
comments and illustration of this method are presented in Sec.
\ref{WRL}.

In absence of vector potential (which means $\bpi=\op$ in
Eq.~(\ref{green2}) or $\Theta=0$ in Eq.~(\ref{greenB1})), the
method and result are similar to those presented earlier by
Gesztesy {\it et al.}\cite{Gesztesy84} and more recently by
Brouder {\it et al.}\cite{Brouder96a,Brouder96b} for the non
magnetic case, although the form we give here is simpler than the
one given by these authors. The generalization to magnetic cases
suggested in Refs.\onlinecite{Brouder96a,Brouder96b} is not
correct because the authors, who have started from $H^B$, have
omitted the factor $\beta$ in the expression of the exchange
coupling $-\mu_B \beta (\bsig\cdot {\bf B}_{\rm eff})$.

In the limit of low electron energies, it is useful to have a
semi-relativistic expression. In order to get it, we perform an
expansion of the operator D in powers of $1/c$ which allows to
write the Dirac Green's function as a series
\begin{eqnarray}\label{green3}
  G(z)=\sum_{n=0}^{\infty} G^{(n)}(z) ,
\end{eqnarray}
where $G^{(n)}(z)$ is the term of order $1/c^n$. The successive
terms are
\wide{m}{
\begin{eqnarray}\label{greennr}
     G^{(0)}(z)&=&\left(\begin{array}{cc}\tG(z)&0\\0&0\end{array}\right) , \\
     G^{(2k+1)}(z) &=&
     \left(\begin{array}{cc}
     0 & \tG(z)\frac{\bssig\cdot\bspi}{2mc}\left(-\frac{(z-V)}{2mc^2}
     Q(z)\right)^k\\
     \left(-Q(z)\frac{(z-V)}{2mc^2}\right)^k
     \frac{\bssig\cdot\bspi}{2mc}\tG(z) & 0
     \end{array}\right) , \\
     G^{(2k+2)}(z) &=&
     \left(\begin{array}{cc}
     -\tG(z)\frac{\bssig\cdot\bspi}{2mc} (z-V)\left(-Q(z)
     \frac{(z-V)}{2mc^2}\right)^k
     \frac{\bssig\cdot\bspi}{2mc}\tG(z) & 0\\
     0 & \frac{1}{2mc^2} \left(-Q(z)\frac{(z-V)}{2mc^2}\right)^k Q(z)
     \end{array}\right) ,
\end{eqnarray}
}
for $k \ge 0$. We remark that odd terms in the expansion of the
Green's function in powers of $1/c$ are odd matrices whereas even
terms are even matrices. Corresponding expressions for the Dirac
Green's function expansion associated with $H^B$ are given in
Appendix A.

One advantage of the expression (\ref{green3}) is that we can
directly identify and calculate the terms which gives rise to
particular effect according to the order of this effect with
$1/c$. Consider for example the magnetic anisotropies. In the
case of a system with uniaxial anisotropy, the anisotropy energy
is quadratic in the spin-orbit coupling $\lambda_{so}$ (i.e., of
order $1/c^4$). As the anisotropy energy is linear with respect
to the Green's function, we have to consider the term $G^{(4)}$.
For cubic anisotropy, the anisotropy energy is of order
$\lambda_{so}^4$ (i.e., of order $1/c^8$), then we have to
consider the term $G^{(8)}$. For galvanomagnetic and
magneto-optical effects, one needs to calculate the conductivity
tensor. The latter is expressed as a product of two Green's
functions and two velocity operators, which are $c\balph$ in the
relativistic theory. For the effects which are linear in
spin-orbit coupling, (i.e., of order $1/c^2$), such as the
anomalous Hall effect, and the Kerr and Faraday magneto-optical
effects, one needs to calculate all terms up to $G^{(4)}$. For
the effects that are quadratic in spin-orbit coupling (i.e., of
order $1/c^4$), such as the AMR or the magnetic birefringence,
one needs all terms up to $G^{(6)}$. Note that the usual (i.e.,
non-relativistic) conductivity is obtained from the terms up to
$G^{(2)}$.

\section{Weak relativistic limit}\label{WRL}
\subsection{Effective Hamiltonian}
We turn now our attention to the effective Hamiltonian which can
be obtained through different ways such as for example the
elimination of the lower components of the wave function or the
Foldy-Wouthuysen transformation which requires a succession of
canonical transformations\cite{Foldy50}. As we have the explicit
expression of the Dirac Green's function in term of $(2 \times
2)$ matrices, we do not need to used such methods to separate the
particles from the antiparticles. Actually, a block
diagonalization of (\ref{green2}) allows to cancels the terms
which couple the upper and lower components and then to extract
the effective Hamiltonian. Let us start with the non-relativistic
limit ($c \rightarrow \infty$) which is obtained in a transparent
manner. Indeed, in this limit, the matrix elements of $G=G^{(0)}$
are different from zero only in the left upper part (see
Eq.~(14)). Then the separation between particles and antiparticles
is naturally made: the Green's function which described the
particle are directly done by $\tG$ and the effective Hamiltonian
is $\tH^A$. In the general case (arbitrary value of c), the
separation between particles and antiparticles can be exactly
made only for free electrons. For particles in a potential, like
in our case, a development in powers of $1/c$ has to be performed
which means a restriction to the weak-relativistic limit. Indeed,
the block diagonalization of the Green's function is only
possible if we cut the expansion (\ref{green3}). Such limitation
is a common characteristic for all methods\cite{Strange98}. Since
we restrict our calculations to the lowest order relativistic
corrections to the Hamiltonian, we have to cut the expansion
(\ref{green3}) of the Green's function after the second order
with $1/c$:
\begin{equation}\label{greengamma}
    G(z) \approx
    \left(\begin{array}{cc}
    \tG(z)-\tG(z)
    \frac{\bssig\cdot\bspi}{2mc}(z-V)
    \frac{\bssig\cdot\bspi}{2mc}\tG(z) & \tG(z)\frac{\bssig\cdot\bspi}{2mc} \\
    \frac{\bssig\cdot\bspi}{2mc}\tG(z) & \frac{1}{2mc^2}Q(z)
    \end{array}\right).\no
\end{equation}
The block diagonalization of $G$ corresponds to the change of
basis $M^{-1}GM$ where the $(4 \times 4)$ unitary matrix $M$ is
given by:
\begin{eqnarray}\label{basis2}
   M=\left(\begin{array}{c}
   1+\frac{(\bssig\cdot\bspi)^2}{4m^2c^2}
   \end{array}\right)^{-\frac{1}{2}}
   \left(\begin{array}{cc}
   1 & -\frac{\bssig\cdot\bspi}{2mc}\\
   \frac{\bssig\cdot\bspi}{2mc} & 1
   \end{array}\right),
\end{eqnarray}
and leads to the block diagonal Green's function:
\begin{eqnarray}\label{basis1}
   M^{-1}G(z)M=
   \left(\begin{array}{cc}
   g_+(z) & 0\\
   0 & g_-(z)
   \end{array}\right).
\end{eqnarray}
By means of this transformation, we have achieved a decoupling
between the particles and the antiparticules:  the $(2\times 2)$
matrix $g_+$ describes the particles whereas the  $(2\times 2)$
matrix $g_-$ describes the antiparticles. We get:
\begin{eqnarray}\label{gpar}
    g_+(z)&=&\tG(z)+\frac{(\bsig\cdot\bpi)^2}{8m^2c^2}\tG(z)+\tG(z)\frac{(\bsig\cdot\bpi)^2}{8m^2c^2} \no \\
    & &-\tG(z)\frac{\bsig\cdot\bpi}{2mc}(z-V)\frac{\bsig\cdot\bpi}{2mc}\tG(z),
\end{eqnarray}
and
\begin{eqnarray}\label{gantipar}
    g_-(z)=\frac{1}{2mc^2}.
\end{eqnarray}
Since now, we restrict our study to the particles. Using the
relations $\tG z = 1 + \tG \tH^A = 1 + \tH^A \tG$, we can
transform the last term in (\ref{gpar}) and write $g_+$ under the
form $\tG+\tG H_{rc}^A\tG$ where:
\begin{eqnarray}\label{hcorr}
  H_{rc}^A&=&-\frac{(\bsig\cdot\bpi)^4}{8m^3c^2}
  +\frac{1}{4m^2c^2}(\bsig\cdot\bpi)V(\bsig\cdot\bpi) \no \\
  & &-\frac{(\bsig\cdot\bpi)^2}{8m^2c^2}V
  -V\frac{(\bsig\cdot\bpi)^2}{8m^2c^2}.
\end{eqnarray}
This expression corresponds to the relativistic corrections of
order $1/c^2$ to the non-relativistic Hamiltonian $\tH^A$. Thus,
the effective Hamiltonian, i.e. Pauli Hamiltonian, is
$H_{\eff}^A=\tH^A+H_{rc}^A$ (it can also be derived from a block
diagonalization of $H^A$). This result is exactly similar to that
given in the literature in the case of a single particle Dirac
equation (see for example Ref. \onlinecite{Messiah62}). In order
to get a more usual expression of $H_{rc}^A$, we have to perform
some transformations which are detailed in Appendix B. In the
case of an uniform effective magnetic field and neglecting
orbital magnetism, $H^A_{rc}$ reduces to (see Eq.~(\ref{HrcAuni})
in Appendix B):
\begin{eqnarray}\label{hrcAuni}
  H_{rc}^A&=&-\frac{p^4}{8m^3c^2}
    +\frac{\hbar^2}{8m^2c^2}\Delta V+\frac{\hbar}{4m^2c^2}\bsig\cdot(\bnab V\times\bpi)\no \\
    &&+\frac{\mu_B}{2m^2c^2}p^2(\bsig\cdot {\bf B}_{\eff}),
\end{eqnarray}
We obtain the usual relativistic corrections  (relativistic mass
correction, Darwin term and spin-orbit coupling) plus an
additional contribution due to the presence of the exchange
coupling. For a non-uniform effective magnetic field, which is the
case in realistic problems, further relativistic corrections are
obtained (see Eq.~(\ref{HrcAdia})). Similar calculations,
presented in Appendix C, have been done starting from $H^B$ by
performing a block diagonalization of the Dirac Green's function
calculated in Appendix A. We can now compare the relativistic
corrections $H_{rc}^A$ and $H_{rc}^B$ obtained in the two
descriptions in the case of a uniform effective magnetic field.
Comparing (\ref{hrcAuni}) and (\ref{hrcBuni}), we observe two
differences: one in the spin-orbit coupling because the
non-relativistic velocity $\tilde{{\bf v}}$ is equal to $\bpi/m$
in the first description whereas it is equal to $\op/m$ in the
second description; and an other one in the relativistic
corrections $H_{rxc}$ to the exchange coupling, given by the last
term in Eqs.~(\ref{hrcAuni}) and (\ref{hrcBuni}), respectively.
Actually, from (\ref{hrcAuni}) we have:
\begin{eqnarray}\label{hxcA}
  H_{rxc}^A\equiv\frac{\mu_B}{2m^2c^2}p^2(\bsig\cdot{\bf B}_{\eff}),
\end{eqnarray}
whereas from (\ref{hrcBuni}) we have:
\begin{eqnarray}\label{hxcB}
  H_{rxc}^B\equiv\frac{\mu_B}{2m^2c^2}(\bsig\cdot\op)(\op\cdot{\bf B}_{\eff}).
\end{eqnarray}
What is problematic is that $H_{rxc}^A$ and $H_{rxc}^B$ couple
the spin and momentum in a quite different manner. Comments and
consequences of this difference are presented in the section IV.

\subsection{Effective velocity operator}
To complete this study, we want to comment briefly on the
velocity operator which is by definition:
\begin{eqnarray}\label{vdef}
  {\bf v}=\frac{1}{i\hbar}[{\bf r},H]
  =\frac{1}{i\hbar}({\bf r}H-H{\bf r})
  =\frac{\partial H}{\partial{\op}},
\end{eqnarray}
where ${\bf r}$ is the position operator. When we report the
expression of the Hamiltonians $H^A$ or $H^B$, we get the simple
form:
\begin{eqnarray}\label{v}
  {\bf v}=c\,\balph=\left(\begin{array}{cc}
    0 & c\,\bsig \\
    c\,\bsig & 0
  \end{array}\right).
\end{eqnarray}
The effective velocity can be obtained from Eq.~(\ref{v}) by the
change of basis $M^{-1}{\bf v}M$ but in order to get the
corrections of order $1/c^2$ to the velocity, it would be
necessary to expand $M$ up to the order $1/c^4$ which is
cumbersome. It can also be obtained form $H_{\eff}^A$ using :
\begin{eqnarray}\label{veffdef}
  {\bf v}_{\eff}^A&=&\frac{1}{i\hbar}[{\bf r},H_{\eff}^A] \no \\
  &=&\frac{1}{i\hbar}[{\bf r},\tH^A]+\frac{1}{i\hbar}[{\bf r},H_{rc}^A]
  \equiv{\bf\tilde{v}}^A+{\bf v}_{rc}^A,
\end{eqnarray}
where ${\bf\tilde{v}}^A$ is the non-relativistic velocity and
${\bf v }_{rc}^A$ the relativistic corrections of order $1/c^2$ to
the velocity. From the expression of $\tH^A$, we get
${\bf\tilde{v}}^A=\bpi/m$ and from (\ref{HrcAdia}) where
diamagnetic terms are neglected, we get:
\begin{eqnarray}
  {\bf v}_{rc}^A&=&-\frac{p^2\op}{2m^3c^2}
  +\frac{\hbar}{4m^2c^2}(\bsig\times\bnab V) \no \\
  &&+\frac{\mu_B}{2m^2c^2}(\op(\bsig\cdot {\bf B}_{\eff})+(\bsig\cdot {\bf B}_{\eff})\op) \no \\
  &&+\frac{e}{4m^3c^2}(p^2{\bf A}+{\bf A}p^2+\op(\op\cdot{\bf A}+{\bf A}\cdot\op) \no \\
  &&+(\op\cdot{\bf A}+{\bf A}\cdot\op)\op).
\end{eqnarray}
In the case of a uniform effective magnetic field and in absence
of orbital magnetism, it reduces to:
\begin{eqnarray}\label{veff}
  {\bf v}_{rc}^A=-\frac{p^2\op}{2m^3c^2}
  +\frac{\hbar}{4m^2c^2}(\bsig\times\bnab V) \no \\
  +\frac{\mu_B}{m^2c^2}\op(\bsig\cdot {\bf B}_{\eff}).
\end{eqnarray}
The first term is the contribution which comes from the
relativistic mass correction. The second term, the so-called
anomalous velocity, results from the spin-orbit coupling and can
play an important role, for example, it leads to the side-jump
mechanism in the anomalous Hall effect. The last term is due to
the presence of the exchange coupling and has no specific name.
Its symmetry is also different in comparison to the relativistic
corrections to the velocity obtained starting from $H^B$ (see
Eq.~(\ref{veffBuni}) in Appendix D).

\section{Discussion}\label{Conc}
This study gives some clarification concerning the assumptions
made when one replaces the Hamiltonian $H^A$ by the Hamiltonian
$H^B$. Even if the consequences of such approximation are not
fully known as K\"ubler justly notices \cite{Kuebler00}; it was
generally believed that the transformation from $H^A$ to $H^B$
neglects only orbital magnetism effects. However, the calculations
of the weak-relativistic limits of $H^A$ and $H^B$ made in Sec.
III, reveal an additional difference which corresponds to a
different symmetry of the relativistic corrections to the exchange
coupling: whereas $H^A_{rxc}$ (see Eq.~(\ref{hxcA})) is isotropic
with respect to the direction of the momentum, $H^B_{rxc}$ (see
Eq.~(\ref{hxcB})) is anisotropic because its amplitude depends,
through the scalar product $(\op\cdot{\bf B}_{\eff})$, of the
angle between the directions of the momentum and the effective
magnetic field. Thus, the use of $H^B$ can lead to anisotropic
effects, such as the magnetocrystalline anisotropy or the AMR,
which differ from those obtained from the exact Kohn-Sham-Dirac
Hamiltonian $H^A$.

Let us first consider the magnetocrystalline anisotropy of a
uniaxial system (e.g., a material with an hexagonal lattice, or
an ultrathin film). If we start from $H^A$, the magnetic
anisotropy arises only as a second order perturbation due to the
spin-orbit coupling, so that it is of order $1/c^4$. In contrast,
if we use $H^B$, it is easy to see that the relativistic
correction of the exchange interaction, $H^B_{rxc}$, gives rise
to an additional contribution to the magnetocrystalline
anisotropy, already in the first order of perturbation, i.e., of
order $1/c^2$, which is unphysical.

Although the details are somewhat more complicated, a similar
result is obtained when considering the AMR of a cubic system:
starting from $H^A$, the AMR arises as a second order
perturbation due to the spin-orbit coupling, i.e., it is of order
$1/c^4$, whereas starting from $H^B$, an additional AMR term of
order $1/c^2$ arises as a first order perturbation due to
$H^B_{rxc}$.

\begin{figure}
\centering \epsfig{file=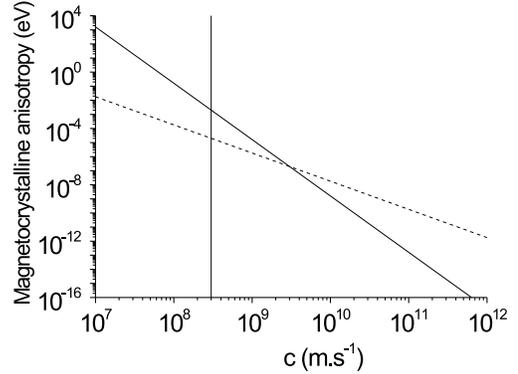} \vspace{0.25cm} \caption{This
figure shows the typical variation of the magneto\-crystalline
anisotropy of a uniaxial ferromagnet as a function of $c$. The
physical value of $c$ is indicated by the vertical line. The
solid curve is the correct value $K^A$ obtained from $H^A$,
whereas the dashed line represents the spurious contribution
$K^B$ obtained from $H^B$.}
\end{figure}

In the limit of $c\to \infty$, these spurious terms of order
$1/c^2$ dominate over the correct terms of order $1/c^4$, which
is of course unacceptable. As all {\it ab initio} calculations of
magnetocrystalline anisotropy\cite{Ujfalussy96,Razee97} starting
for the Dirac equation rely on the approximation (\ref{hrel}),
there validity can be questioned {\it a priori}. Let us make a
simple estimate of the orders of magnitude for the physical value
of $c$. Starting from $H^A$, the magnetocrystalline anisotropy
is\cite{Bruno93}
\begin{equation}
K^A \approx \frac{\lambda_{so}^{2}}{W} \sim \frac{1}{c^4} ,
\end{equation}
where $\lambda_{so}$ is the spin-orbit constant and $W$ the
band-width. Starting from $H^B$ one other hand, it is quite easy
to calculate the magnetocrystalline anisotropy due to $H^B_{rxc}$:
\begin{equation}
  K^B\approx \frac{\varepsilon_F\Delta_{ex}}{2mc^2} .
\end{equation}
where $\varepsilon_F$ is the Fermi level and $\Delta_{ex}$ the
exchange splitting. For a transition metal ferromagnet, by taking
typical values $\lambda_{so} \approx 0.1$~eV and $W\approx 5$~eV,
one obtains $K^A \approx 2\times 10^{-3}$~eV. Taking
$\varepsilon_F\approx 10$~eV, $\Delta_{ex}\approx 2$~eV, and
$mc^{2}\approx 500$~keV, we obtain $K^B\approx 2\times
10^{-5}$~eV. Therefore, in spite of the fact that $K^B\gg K^A$ in
the limit $c\rightarrow\infty$, we find that $K^B\ll K^A$ for the
the physical value $c\approx 3\times 10^8$~m.s$^{-1}$. This
result is visualized in Fig.~1. Therefore the quantitative
results of first-principles calculations of the
magnetocrystalline anisotropy based on the approximate
Kohn-Sham-Dirac Hamiltonian $H^B$ should not be perturbed in a
significant manner by the spurious contribution of order
$1/c^{2}$. In spite of this fortunate circumstance, it would be
desirable to develop a more satisfactory theoretical approach
which is free from unphysical spurious contributions.

For a non-uniform effective magnetic field, additional
relativistic corrections are obtained (see Eqs.~(\ref{HrcAdia})
and (\ref{hrcB})). The qualitative importance of these corrections
remains to be investigated.

To conclude, we want to underline two aspects which represent the
important results of this work. On the one hand, we have derived a
convenient form of the Dirac Green's function which is valid for
any value of $c$ and has the advantage to be express in term of
$(2 \times 2)$ matrices, in particular the non-relativistic Green
function. This is a quite general form which can be used to treat
different problems. We have applied it as an alternative way to
separate particles from anti-particles in order to extract the
effective Hamiltonian and effective velocity in the
weak-relativistic limit. On the other hand, we have performed a
detailed comparison in the weak-relativistic limit between the
Kohn-Sham-Dirac equation and its approximate form containing the
exchange coupling. This study has revealed a difference of
symmetry in the relativistic corrections to the exchange coupling
which can leads to artificial anisotropic effects. A qualitative
estimation has shown that, in the case of transition metals, this
difference is not significant.
\end{multicols}
\appendix
\begin{multicols}{2}
\section{DIRAC GREEN'S FUNCTION IN PRESENCE OF AN EXCHANGE COUPLING}
In this appendix, we summarize the derivation of the Dirac
Green's function starting from $H^B$ given by (\ref{hrel}). It
follows the same steps than starting from $H^A$ but involves
different matrices and leads to a different expression of the
final Dirac Green's function. To simplify the notations, we
introduce $\Theta = -\mu_B\left(\bsig\cdot\Beff\right)$, then:
\begin{eqnarray}
  H^B=\left(\begin{array}{cc}
  V+\Theta & c\left(\bsig\cdot\op\right) \\
  c\left(\bsig\cdot\op\right) & V-2mc^2-\Theta
  \end{array}\right) .
\end{eqnarray}
The Dirac Green's function $G(z)\equiv(z-H^B)^{-1}$ can then be
written as:
\begin{eqnarray}\label{manipB}
  G(z)
  &=&\left(1+{\mathcal A}^{-1}(z)
  \left(\begin{array}{cc}
  -V-\Theta & 0 \\
  0 & z-V+\Theta
  \end{array}\right)\right)^{-1} \no \\
  &&\times {\mathcal A}^{-1}(z).
\end{eqnarray}
${\mathcal A}^{-1}$ is given by (\ref{gint2}) when we replace
$\bpi$ by $\op$ and define $\tG_0(z)\equiv(z-p^2/2m)^{-1}$. We
perform the direct inversion of the matrices which appear in
(\ref{manipB}). It leads to the final expression:
\wide{m}{
\begin{eqnarray}\label{greenB1}
  G(z)=\left(\begin{array}{cc}
  \tG(z)-\tG(z)\frac{\bssig\cdot\op}{2mc}
        (z-V+\Theta)D(z)\frac{\bssig\cdot\op}{2mc}\tG(z)
  & \tG(z)\frac{\bssig\cdot\op}{2mc} Q^{-1}(z)D(z)Q(z)\\
  D(z)\frac{\bssig\cdot\op}{2mc}\tG(z)
  & \frac{1}{2mc^2}D(z)Q(z)
  \end{array}\right),
\end{eqnarray}
}
where $\tG$ is the ($2\times 2$) non-relativistic Green's
function associated with the ($2\times 2$) non-relativistic
Hamiltonian $\tH^B = p^2 / 2m + V + \Theta$, and the operators
$D$ and $Q$ are given by:
\begin{eqnarray}
D(z)&=&\left(1+Q(z)\frac{(z-V+\Theta)}{2mc^2}\right)^{-1}, \\
Q(z)&=&1+\frac{(\bsig\cdot\op)\tG(z)(\bsig\cdot\op)}{2m}.
\end{eqnarray}
A semi-relativistic expansion $G(z) =
\sum_{n=0}^{\infty}G^{(n)}(z)$ can also be given. The successive
terms are
\wide{m}{
\begin{eqnarray}\label{greenB2}
     G^{(0)}(z)&=&\left(\begin{array}{cc}\tG(z)&0\\0&0\end{array}\right) , \\
     G^{(2k+1)}(z) &=&
     \left(\begin{array}{cc}
     0 & \tG(z)\frac{\bssig\cdot\op}{2mc}\left(-\frac{(z-V+\Theta)}{2mc^2}
     Q(z)\right)^k\\
     \left(-Q(z)\frac{(z-V+\Theta)}{2mc^2}\right)^k
     \frac{\bssig\cdot\op}{2mc}\tG(z) & 0
     \end{array}\right) , \\
     G^{(2k+2)}(z) &=&
     \left(\begin{array}{cc}
     -\tG(z)\frac{\bssig\cdot\op}{2mc} (z-V+\Theta)\left(-Q(z)
     \frac{(z-V+\Theta)}{2mc^2}\right)^k
     \frac{\bssig\cdot\op}{2mc}\tG(z) & 0\\
     0 & \frac{1}{2mc^2} \left(-Q(z)\frac{(z-V+\Theta)}{2mc^2}\right)^k Q(z)
     \end{array}\right) ,
\end{eqnarray}
for $k \ge 0$.\\
}
\section{EFFECTIVE HAMILTONIAN IN PRESENCE OF A POTENTIAL VECTOR}
Transformations of $\tH^A$ and $ H_{rc}^A$ allow to get a more
usual expression of the effective Hamiltonian $H_{\eff}^A$.
Neglecting diamagnetic terms, we have:
\begin{eqnarray}
    \tH^A=\frac{(\bsig\cdot\bpi)^2}{2m}+V&=&\frac{p^2}{2m}+V-\mu_B(\bsig\cdot {\bf B}_{\eff}) \no \\
    &&-\frac{e}{2m}(\op\cdot{\bf A}+{\bf A}\cdot\op),
\end{eqnarray}
where we have used the identities:
\begin{eqnarray}
  &&\left(\bsig\cdot\bpi\right)\left(\bsig\cdot\bpi\right)
  =\bpi \cdot\bpi+i\bsig\cdot\left(\bpi \times\bpi\right), \no \\
  &&\op\times{\bf A}+{\bf A}\times\op=-i\hbar{\bf B}_{\eff}.
\end{eqnarray}
For a uniform effective field, $\tH^A$ reduces to:
\begin{eqnarray}
    \tH^A=\frac{p^2}{2m}+V-\mu_B((\bsig+{\bf L})\cdot {\bf B}_{\eff}),
\end{eqnarray}
where ${\bf L}={\bf r}\times\op$ is the orbital momentum. In
absence of orbital magnetism, this last expression is identical to
$\tH^B$. We turn now our attention to the relativistic
corrections $H_{rc}^A$. After transformation of (\ref{hcorr}) and
neglecting the diamagnetic terms, we can rewrite $H_{rc}^A$ as:
\begin{eqnarray}\label{HrcAdia}
    H_{rc}^A&=&-\frac{p^4}{8m^3c^2}+\frac{\hbar^2}{8m^2c^2}\Delta V+\frac{\hbar}{4m^2c^2}\bsig\cdot(\bnab V\times\bpi) \no \\
    &&+\frac{\mu_B}{4m^2c^2}\left(p^2(\bsig\cdot {\bf B}_{\eff})+(\bsig\cdot {\bf B}_{\eff})p^2\right) \no \\
    &&+\frac{e}{8m^2c^2}\left(\frac{p^2}{m}(\op\cdot{\bf A}+{\bf A}\cdot\op)
    +(\op\cdot{\bf A}+{\bf A}\cdot\op)\frac{p^2}{m}\right. \no \\
    &&+\left.({\bf A}\cdot\op-\op\cdot{\bf A})V-V({\bf A}\cdot\op-\op\cdot{\bf A})\right),
\end{eqnarray}
where we have used the following identities:
\begin{eqnarray}\label{ident}
  &&\left(\bsig\cdot\bpi\right)V\left(\bsig\cdot\bpi\right)
  =\bpi V\cdot\bpi+i\bsig\cdot\left(\bpi V\times\bpi\right), \no \\
  &&2\op\,V\cdot\op=p^2V+Vp^2+\hbar^2\Delta V,
\end{eqnarray}
and replaced the momentum operator $\op$ as well as the operator
$p^2$, when they act only on the potential, respectively by the
gradient $\bnab=-\op/i\hbar$ and by the Laplacian
$\Delta=-p^2/\hbar^2$. In the case of a uniform effective
magnetic field, $H_{rc}^A$ reduces to:
\begin{eqnarray}\label{HrcAuni}
  H_{rc}^A&=&-\frac{p^4}{8m^3c^2}
  +\frac{\hbar^2}{8m^2c^2}\Delta V+\frac{\hbar}{4m^2c^2}\bsig\cdot(\bnab V\times\bpi)\no \\
  &&+\frac{\mu_B}{2m^2c^2}p^2((\bsig+{\bf L})\cdot {\bf B}_{\eff}).
\end{eqnarray}

\section{EFFECTIVE HAMILTONIAN IN PRESENCE OF AN EXCHANGE COUPLING}
In this appendix, we summarize the derivation of the effective
Hamiltonian starting from the Dirac Green's function
(\ref{greenB1}). We cut the expansion after the second order with
$1/c$, then:
\wide{m}{
\begin{equation}\label{greenB3}
    G(z) \approx
    \left(\begin{array}{cc}
    \tG(z)-\tG(z)
    \frac{\bssig\cdot\op}{2mc}(z-V+\Theta)
    \frac{\bssig\cdot\op}{2mc}\tG(z) & \tG(z)\frac{\bssig\cdot\op}{2mc} \\
    \frac{\bssig\cdot\op}{2mc}\tG(z) & \frac{1}{2mc^2}Q(z)
    \end{array}\right).
\end{equation}
}
The block diagonalization $M^{-1}GM$ is
obtained with:
\begin{eqnarray}\label{basisB}
   M=\left(\begin{array}{c}
   1+\frac{p^2}{4m^2c^2}
   \end{array}\right)^{-\frac{1}{2}}
   \left(\begin{array}{cc}
   1 & -\frac{\bssig\cdot\op}{2mc}\\
   \frac{\bssig\cdot\op}{2mc} & 1
   \end{array}\right),
\end{eqnarray}
and leads to:
\begin{eqnarray}\label{gparB}
    g_+(z)&=&\tG(z)+\frac{p^2}{8m^2c^2}\tG(z)+\tG(z)\frac{p^2}{8m^2c^2} \no \\
    & &-\tG(z)\frac{\bsig\cdot\op}{2mc}(z-V+\Theta)\frac{\bsig\cdot\op}{2mc}\tG(z),
\end{eqnarray}
which can be written under the form $\tG+\tG H_{rc}^B\tG$ where:
\begin{eqnarray}\label{hcorrB}
  H_{rc}^B&=&-\frac{p^4}{8m^3c^2}
  +\frac{1}{4m^2c^2}\left(\bsig\cdot\op\right)\left
  (V-\Theta\right)\left(\bsig\cdot\op\right) \no \\
  & &-\frac{p^2}{8m^2c^2}\left(V+\Theta\right)
  -\left(V+\Theta\right)\frac{p^2}{8m^2c^2}.
\end{eqnarray}
In order to get a more usual expression of $H_{rc}^B$, we made
some transformations of (\ref{hcorrB}) using identities
(\ref{ident}) where we have replaced $\bpi$ by $\op$ and the
relation:
\begin{eqnarray}\label{identB}
  &\left(\bsig\cdot\op\right)\left(\bsig\cdot\Beff\right)\left(\bsig\cdot\op\right)=& \no \\
  &-\op\left(\bsig\cdot\Beff\right)\cdot\op
  +\hbar\left(\bnab\times\Beff\right)\cdot\op& \no \\
  &+\left(\bsig\cdot\op\right)\left(\Beff\cdot\op\right)
  +\left(\op\cdot\Beff\right)\left(\bsig\cdot\op\right).&
\end{eqnarray}
Finally, we get:
\begin{eqnarray}\label{hrcB}
  & &H_{rc}^B=-\frac{p^4}{8m^3c^2}
  +\frac{\hbar^2}{8m^2c^2}\Delta\Big(V-\mu_B(\bsig\cdot\Beff)\Big) \no \\
  & &+\frac{\hbar}{4m^2c^2}\bsig\cdot(\bnab V\times\op)
  +\frac{\hbar\mu_B}{4m^2c^2}(\bnab\times\Beff)\cdot\op \no \\
  & &+\frac{\mu_B}{4m^2c^2}\Big((\bsig\cdot\op)(\Beff\cdot\op)
  +(\op\cdot\Beff)(\bsig\cdot\op)\Big).
\end{eqnarray}
This expression of the effective Hamiltonian in presence of the
exchange coupling is equivalent to the one obtained by Kraft {\it
et al.}\cite{Kraft95}. In the case of a uniform effective
magnetic field, it reduces to:
\begin{eqnarray}\label{hrcBuni}
  H_{rc}^B&=&-\frac{p^4}{8m^3c^2}
  +\frac{\hbar^2}{8m^2c^2}\Delta V
  +\frac{\hbar}{4m^2c^2}\bsig\cdot(\bnab V\times\op) \no \\
  & &+\frac{\mu_B}{2m^2c^2}(\bsig\cdot\op)(\Beff\cdot\op).
\end{eqnarray}

\section{EFFECTIVE VELOCITY OPERATOR IN PRESENCE OF AN EXCHANGE COUPLING}
In this appendix, we summarize the derivation of the effective
velocity starting from $H^B$. From the expression of
$\tH^B=p^2/2m+V+\Theta$, we get ${\bf\tilde{v}}^B=\op/m$ and from
the expression (\ref{hrcB}) of $H_{rc}^B$, we get:
\begin{eqnarray}\label{veffB}
  {\bf v}_{rc}^B&=&-\frac{p^2\op}{2m^3c^2}
  +\frac{\hbar}{4m^2c^2}(\bsig\times\bnab V)+\frac{\hbar\mu_B}{4m^2c^2}(\bnab\times\Beff) \no \\
  & &+\frac{\mu_B}{4m^2c^2}\Big(\bsig(\Beff\cdot\op)+(\op\cdot\Beff)\bsig\no \\
  & &+(\bsig\cdot\op)\Beff+\Beff(\bsig\cdot\op)\Big).
\end{eqnarray}
This expression differs from the result obtained by Kraft {\it et
al.}\cite{Kraft95} on two points: they get a wrong coefficient (a
factor 2 missing) for the contribution of the relativistic mass
correction and they obtain a contribution from the Darwin term
which should not appear because the commutator $[{\bf r},\Delta
V]$ is equal to zero. In the case of a uniform effective magnetic
field, (\ref{veffB}) reduces to:
\begin{eqnarray}\label{veffBuni}
  {\bf v}_{rc}^B&=&-\frac{p^2\op}{2m^3c^2}
  +\frac{\hbar}{4m^2c^2}(\bsig\times\bnab V) \no \\
  & &+\frac{\mu_B}{2m^2c^2}\Big(\bsig(\Beff\cdot\op)+\Beff(\bsig\cdot\op)\Big).
\end{eqnarray}

\end{multicols}

\end{document}